%% file: main.tex
\documentclass[sigconf]{acmart}

\input{constants/imports.tex}

\AtBeginDocument{%
  }

\setcopyright{none}
\acmConference{}{}{}
\acmDOI{}
\acmISBN{}
\begin{document}

\title{\dotrag{}: Retrieval-Time Reasoning Along Relational Paths}

\author{Larnell Moore}\affiliation{\institution{University of Michigan}\country{USA}}
\author{Naihao Deng}\affiliation{\institution{University of Michigan}\country{USA}}
\author{Rada Mihalcea}\affiliation{\institution{University of Michigan}\country{USA}}
\author{Farnaz Jahanbakhsh}\affiliation{\institution{University of Michigan}\country{USA}}
\input{sections/abstract}

\ccsdesc[500]{Information systems~Language models}
\ccsdesc[500]{Information systems~Specialized information retrieval}

\keywords{Graph Retrieval-Augmented Generation, graph retrieval, relational reasoning, knowledge graphs, large language models}

\maketitle
\input{sections/intro}

\input{sections/related_works}

\input{sections/method}
\input{sections/experiments}

\bibliographystyle{ACM-Reference-Format}
\bibliography{references}

\input{sections/appendix}

\end{document}

%% file: constants/imports.tex
\usepackage[table]{xcolor}

\usepackage{booktabs}

\usepackage{arydshln}

\usepackage{multirow}

\newcommand{\dotrag}{DotRAG}

\usepackage{graphicx}

%% file: sections/abstract.tex
\begin{abstract}
Graph Retrieval-Augmented Generation (GraphRAG) faces significant challenges in complex relational retrieval tasks, particularly in densely connected knowledge graphs where the number of possible paths grows rapidly with reasoning depth. Existing GraphRAG retrieval algorithms either rely on rigid, predefined heuristics to identify relevant graph context or employ costly LLM-based reasoning to guide graph traversal. To address these limitations, we propose \dotrag{}, a lightweight, training-free GraphRAG retrieval algorithm that uses large language model (LLM) reasoning to narrow the search space and recover task-relevant relational paths. During indexing, \dotrag{} constructs an entity-relation knowledge graph, generates a graph-wide overview summarizing its contents, and leverages graph reachability indexing to avoid expensive online graph traversals. At query time, the LLM uses this overview to identify promising graph regions and hypothesizes relationship sequences connecting graph information to the query. The LLM then performs targeted relational searches, using graph constraints to restrict the candidate pool progressively. On MetaQA 2-hop queries, \dotrag{} achieves 97.78\% retrieval recall and 51.27\% F1, compared to the best baseline scores of 34.74\% and 6.33\%, respectively. On STaRK-MAG, \dotrag{} achieves 85.07\% recall on 2-hop queries while demonstrating greater computational efficiency than reasoning-based GraphRAG baselines, with 12.2$\times$ lower retrieval latency than HopRAG.
\end{abstract}

%% file: sections/intro.tex
\section{Introduction}
    Retrieval-Augmented Generation (RAG) grounds large language models (LLMs) to external knowledge  \citep{lewis2020retrieval, karpukhin2020dense}. Recent work has leveraged graphs (GraphRAG) to enhance the retrieval process \citep{han2024graphragsurvey,peng2025graphragsurvey,
li2025grefer,zhu2024structugraphrag}. However, most existing GraphRAG retrieval methods rely on predefined heuristics, typically based on semantic similarity, graph-topological properties such as node degree, or combinations of these signals \citep{he2024gretriever, edge2024graphrag, guo-etal-2025-lightrag, chen2026pathrag}. These heuristics struggle with complex relational queries that require retrieving specific sequences of relations connecting entities \citep{yu2026graphragr1, liu2025hoprag, lau2026catrag, guo-etal-2026-beyond-static}. This challenge becomes more pronounced in densely connected knowledge graphs, where the number of possible relational paths grows combinatorially with reasoning depth, making it increasingly difficult to identify the paths needed to retrieve the correct evidence for a query \citep{peng2025graphragsurvey, liang2025fasttog, du-etal-2025-mixture, wang-etal-2019-incorporating}.

Consequently, heuristic-driven retrieval can introduce two key failure modes in complex relational tasks: \textit{(i) retrieving irrelevant entities and relationships}, which introduces noise into the model's context and complicates downstream reasoning \citep{chen2026pathrag, guo-etal-2025-empowering, huang2026relink}, or \textit{(ii) failing to retrieve necessary evidence}, leaving the model without the information required to answer a query, even when that information exists in the graph \citep{zhan-etal-2026-flowrag, guo-etal-2026-beyond-static, zhu-etal-2025-mitigating}.

To move away from predefined traversal rules, some GraphRAG retrieval algorithms incorporate LLM reasoning directly into retrieval decisions, allowing the model to determine which graph regions to explore, which relationships to follow, and what context to retrieve. However, these approaches face two key limitations \textit{(i) as graph size and connectivity increase}, the number of retrieval decisions can grow substantially, limiting scalability to large knowledge graphs; and \textit{(ii) the high latency of LLM reasoning} makes repeated retrieval decisions computationally expensive and often impractical.

To address these challenges, we introduce \dotrag{}, a lightweight, training-free GraphRAG retrieval algorithm designed to recover task-relevant relational paths from knowledge graphs. During offline indexing, \dotrag{} builds an entity-relation knowledge graph and generates two complementary information sources to support retrieval. First, it generates a global overview that summarizes the graph's contents, informing the LLM at inference time what information the graph contains. Second, it constructs a graph reachability index that records the shortest-hop distances between entities, telling the LLM which types of entities are reachable and at what distances.  At query time, the LLM uses this information to decide which graph regions to explore and how to constrain the search to find task-relevant relational paths progressively. Thus, \dotrag{} keeps the search task-focused, reduces the number of LLM calls needed for graph exploration by grounding retrieval decisions in the knowledge base, and avoids costly online graph traversals by precomputing structural information in the reachability index.

We evaluate \dotrag{} on STaRK\citep{wu2024stark} and MetaQA\citep{MetaQA}, two challenging relational retrieval benchmarks. A common limitation explicitly acknowledged in prior GraphRAG evaluations is the lack of reliable ground truth for directly assessing retrieval effectiveness, leading many studies to rely on downstream answer quality and LLM-as-a-Judge, which can introduce subjective judgments and model-dependent biases \cite{guo-etal-2025-lightrag, chen2026pathrag, edge2024graphrag}. We use these two benchmarks to ensure that all methods are evaluated on the same knowledge graphs, including any existing graph-construction errors, natural-language queries, and ground-truth answer entities. Both contain graphs derived from automatic extraction pipelines from unstructured data and include realistic GraphRAG noise, such as duplicate entities, mislabeled entities, and other extraction errors. This enables reproducible and fair comparisons that isolate retrieval algorithm performance from differences in graph construction and downstream answer generation. 

Our results show that DotRAG substantially improves multi-hop retrieval effectiveness, achieving 97.78\% recall and 51.27\% F1 on MetaQA 2-hop queries and 85.07\% recall on STaRK-MAG 2-hop queries. These retrieval gains translate into stronger downstream answer quality, with DotRAG achieving the highest answer recall and F1 across all evaluated settings, demonstrating its effectiveness in recovering task-relevant relational evidence that LLMs can successfully utilize. In summary, our contributions are:\begin{enumerate}
    \item Our key contribution is we introduce a novel LLM-guided relational retrieval algorithm that performs localized semantic searches directly over graph relationships, progressively constraining the search space rather than allowing irrelevant candidates from the global graph to compete.

    \item We introduce two complementary graph indexing mechanisms: a compressed global graph overview that provides prior knowledge of the graph's contents and relational structure, and a precomputed reachability map that enables targeted retrieval without exhaustive graph exploration or beam search.

    \item We provide interpretable relational retrieval through structured, logically connected traversal paths that expose how retrieved evidence relates to the query, making retrieval decisions traceable and explainable.

    \item We establish a large-scale relational retrieval evaluation on STaRK and MetaQA. Our largest evaluation graph contains 88,815 entities and 486,616 relationships, exceeding the largest graph evaluated in HopRAG by approximately 119\% in nodes and 183\% in edges.

    \item  We demonstrate that \dotrag{} substantially improves multi-hop retrieval effectiveness and downstream answer quality across MetaQA and STaRK, achieving the highest 2-hop retrieval F1 across all evaluated datasets and the highest answer recall and F1 across all evaluated settings.
\end{enumerate}


%% file: sections/related_works.tex
\section{Related Works}

\begin{figure}[t]
    \centering
    \includegraphics[
        width=\linewidth,
    ]{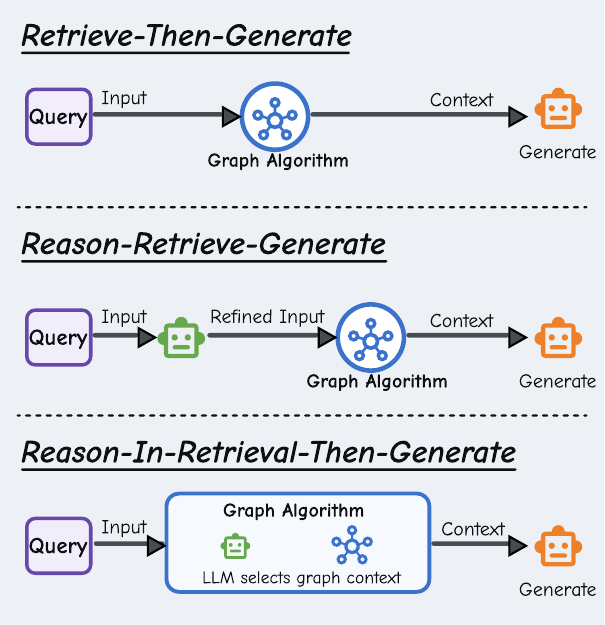}
    \caption{Comparison of retrieval paradigms in GraphRAG.}
    \label{fig:retrieval-paradigms}
\end{figure}

As illustrated in Figure~\ref{fig:retrieval-paradigms}, we distinguish LLM reasoning used during retrieval from reasoning used for final answer generation. Based on the degree of LLM control over retrieval decisions, existing RAG methods can be categorized into three paradigms: Retrieve-Then-Generate, Reason-Retrieve-Generate, and Reason-In-Retrieval-Then-Generate. These paradigms correspond to three levels of LLM control over retrieval, namely no retrieval control, input-level control, and algorithm-level control, respectively.
\paragraph{Retrieve-Then-Generate.}
In this paradigm, retrieval operates without explicit LLM reasoning guiding its decisions. As depicted, the user's original query is passed directly to the retrieval algorithm without invoking an LLM before or during retrieval. The algorithm identifies relevant context from the knowledge source, which is then provided to an LLM for final answer generation. Classic RAG approaches fall under this paradigm \citep{lewis2020retrieval}, commonly employing lexical or semantic retrieval methods such as BM25 \citep{robertson2009probabilistic} and dense vector search \citep{karpukhin2020dense}. While these methods enable efficient and scalable retrieval, a known limitation is that their effectiveness tends to decline as reasoning complexity increases, particularly for multi-hop queries  \citep{trivedi2023interleaving, shen-etal-2025-gear, melo-etal-2025-dont, shi-etal-2024-generate}.

\paragraph{Reason-Retrieve-Generate.}
To address Retrieve-Then-Generate's limitations, prior work uses LLM reasoning to refine the inputs provided to retrieval algorithms. In these approaches, the LLM does not make graph traversal decisions; instead, it reformulates the query and passes it to an existing retrieval algorithm. This may occur in a single step \citep{ye-etal-2023-enhancing, ma-etal-2023-query} or iteratively, where the LLM submits a query, examines the retrieved results, and formulates follow-up queries to retrieve additional information \citep{jiang2023active, yu2026graphragr1, trivedi2023interleaving}. However, the LLM does not control the retrieval algorithm's internal decisions about which evidence is relevant or which graph regions and relationships should be explored. For example, LightRAG~\cite{guo-etal-2025-lightrag} and PathRAG~\cite{chen2026pathrag} use an LLM to transform the original query into keywords, which they then provide as input to their graph retrieval algorithms, while the algorithms independently determine which evidence to retrieve.

\begin{figure*}[t]
    \centering
    \includegraphics[width=\textwidth]{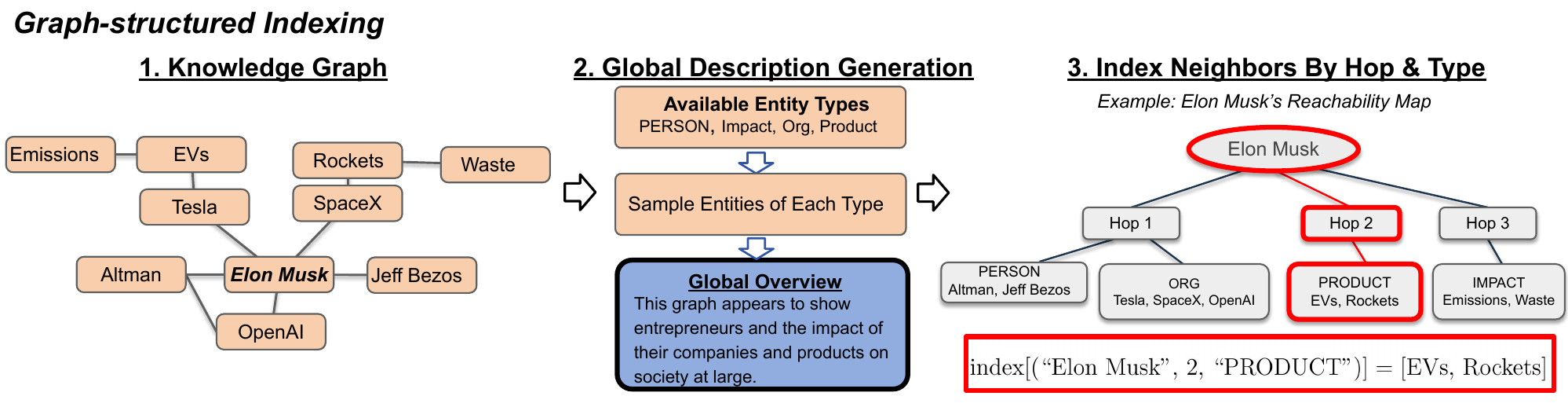}
    \caption{Overview of DotRAG’s offline graph-structured indexing. A global graph overview is generated by summarizing sampled entities from each type, while a reachability map indexes entities by hop distance and type to support efficient, structurally constrained retrieval.}
    \label{fig:indexing}
\end{figure*}

\paragraph{Reason-In-Retrieval-Then-Generate.}
Reason-In-Retrieval approaches incorporate LLM reasoning directly into the graph retrieval process, allowing the LLM to decide which graph regions to explore, which relationships to follow, and which evidence to retrieve \citep{sun2024thinkongraph, han2026roe, hu2025remindrag}. For example, HopRAG~\citep{liu2025hoprag} iteratively uses LLM reasoning during graph traversal to evaluate candidate relationships, select which edges to follow, and progressively retrieve relevant passages across multiple hops through a retrieve-reason-prune mechanism. A key limitation of these approaches is that they often incur substantial computational overhead due to repeated LLM calls during retrieval, increasing latency and cost \citep{liu2025hoprag,liang2025fasttog, lee-etal-2025-hybgrag}.

In contrast, \dotrag{} is a training-free Reason-In-Retrieval-Then-Generate approach that combines LLM-guided retrieval decisions with graph-structural constraints to enable targeted relational path retrieval. By leveraging offline graph indexing, \dotrag{} provides the LLM with prior knowledge of the graph's contents and structure, while progressively restricting semantic search to structurally relevant candidates. This enables informed decisions about which relational paths to explore and when to stop, reducing unnecessary exploration and computational overhead.

%% file: sections/method.tex
\section{Problem Formulation}

Following \citet{guo-etal-2025-lightrag}, we represent knowledge as an undirected entity-relation graph $G=(\mathcal{E},\mathcal{R})$, where $\mathcal{E}$ denotes entities and $\mathcal{R}$ denotes relationships between them. 
Given a natural language query $q$, the goal is to retrieve a set of task-relevant relational paths $\mathcal{P}_q = \{p_1, p_2, \ldots, p_k\}$, where each path $p_i$ represents an ordered sequence of relationships connecting entities in $G$. 

The retrieval objective is to maximize coverage of relevant relational paths while minimizing irrelevant information introduced into the retrieved context. Additionally, we aim to minimize the computational overhead associated with path search and LLM reasoning, enabling the method to scale across graphs of varying sizes.

\section{\dotrag{}}
We introduce \dotrag{} consisting of two main components of offline graph indexing (\S\ref{subsec: graph-structured-indexing}) and a retrieval algorithm (\S\ref{subsec: retrieval-algorithm}) that leverages these components.

\subsection{Graph-Structured Indexing}
\label{subsec: graph-structured-indexing}

We introduce indexing mechanisms that give the retriever prior knowledge of the graph’s contents and relational structure through a compressed global overview of the graph and a precomputed reachability map.

\paragraph{Global Graph Overview.}
To reduce reasoning and exploration during retrieval, we generate a compact overview of the graph's contents and structure. As illustrated in Figure~\ref{fig:indexing}, after graph construction, we sample entities from each predefined type and use an LLM to summarize the patterns observed within each type. The LLM then synthesizes these type-level summaries into a global graph overview, providing prior knowledge to guide retrieval decisions. This process requires only $|\mathcal{T}|+1$ LLM calls, where $\mathcal{T}$ denotes the set of entity types, regardless of graph size.

\begin{figure*}[t]
    \centering
    \includegraphics[width=\textwidth]{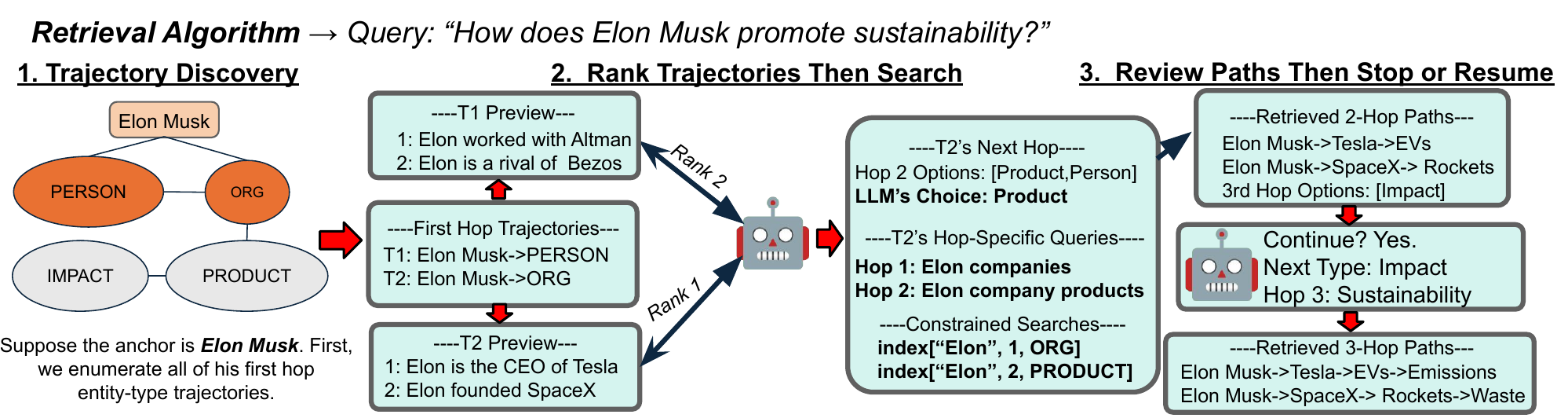}
    \caption{Overview of DotRAG's retrieval algorithm.}
    \label{fig:dotrag-retrieval}
\end{figure*}
\paragraph{Reachability Map.}
Complementing the global overview, we also introduce a reachability map
during graph indexing. The reachability map serves two purposes: enabling
structurally constrained semantic search and avoiding repeated graph
traversal during retrieval.

To construct the map, we perform bounded breadth-first search (BFS) offline
from every entity and record the entities appearing at each hop up to a
fixed radius $H$. For an entity $e \in \mathcal{E}$, we denote the entities
appearing at hop $h$ as $\mathcal{N}_h(e)$. We further organize these sets
according to the predefined entity types:
\[
\mathcal{N}_{h,t}(e)
=
\left\{
v \in \mathcal{N}_h(e)
\mid t \in \tau(v)
\right\},
\]
where $\tau(v)$ denotes the type assignments of entity $v$.

To illustrate the reachability map's usefulness, consider the example in
Figure~\ref{fig:indexing}. The map for \textit{Elon Musk} organizes
neighboring entities by both hop distance and entity type. At the first
hop, it records reachable people, such as \textit{Altman} and
\textit{Jeff Bezos}, and organizations, such as \textit{Tesla},
\textit{SpaceX}, and \textit{OpenAI}. At the second hop, it records
reachable products, including \textit{EVs} and \textit{Rockets}. At the
third hop, it records impacts, including \textit{Emissions} and
\textit{Waste}. Consequently, the lookup
\[
\operatorname{index}[(\textit{Elon Musk},2,\textsc{Product})]
=
\{\textit{EVs},\textit{Rockets}\}
\]
directly returns the product entities located two hops from
\textit{Elon Musk}.

In most GraphRAG approaches, semantic retrieval operates over a global
embedding space. However, global retrieval can introduce noise by allowing
globally dominant but irrelevant entities to enter the retrieved context.
\dotrag{} instead uses the reachability map to first constrain the candidate
set by anchor entity, hop distance, and entity type, and then performs
semantic search only over the resulting set rather than over the entire
graph. In the example above, a search for products related to
\textit{Elon Musk} is therefore restricted to the two-hop product candidates
$\{\textit{EVs},\textit{Rockets}\}$ before they are semantically ranked.

The reachability map also avoids performing a new breadth-first search or
exhaustively enumerating paths during each query. Once constructed, it
supports constant-time lookup of the reachable set associated with an
entity, hop distance, and entity type. Enumerating its contents requires
$O(|\mathcal{N}_{h,t}(e)|)$ time, compared with the
$O(|\mathcal{E}|+|\mathcal{R}|)$ worst-case cost of performing a new BFS
at query time. Thus, the reachability map introduces a deliberate
space--time tradeoff: precomputing bounded reachability for every entity
requires $O(H|\mathcal{E}|^2)$ storage in the worst case, but shifts graph
exploration from inference time to the offline indexing stage.

\subsection{Retrieval Algorithm}
\label{subsec: retrieval-algorithm}
\dotrag{}'s retrieval algorithm proceeds in four key stages:
\emph{trajectory discovery}, \emph{trajectory ranking},
\emph{relational path search}, and \emph{relational path ranking}.
The discovery phase deterministically identifies query-relevant seed
entities and constructs graph-supported retrieval options. The LLM then
ranks these options from most to least promising. Relational path search
extends the highest-ranked options through coordinated semantic searches.
Finally, relational path ranking orders the recovered evidence and selects
the paths that fit within the final token budget.

A key design principle of \dotrag{} is to restrict LLM reasoning to
selecting among graph-supported options and coordinating search. This
avoids the computational and latency overhead of requiring excessive LLM
calls to explore the graph.

\paragraph{Trajectory Discovery.}
To initiate graph search, \dotrag{} begins by performing a global semantic
search over the graph to find seed entities relevant to the query $q$. It
retrieves the top $k$ entities, with $k$ specifying the semantic-search
budget. From these seeds, entities whose names are directly mentioned in
the query are prioritized as candidate anchors. If none satisfy this
condition, \dotrag{} selects the top $k_a$ semantically ranked seeds as
candidate anchors.

For each candidate anchor, \dotrag{} uses the reachability map to enumerate
the entity types available at the first hop. As illustrated in
Figure~\ref{fig:dotrag-retrieval}, the first-hop types reachable from
\textit{Elon Musk} are \textsc{Person} and \textsc{Organization}. These
produce the trajectories
\textit{Elon Musk $\rightarrow$ Person} and
\textit{Elon Musk $\rightarrow$ Organization}.

More generally, the graph-supported first-hop trajectories rooted at
candidate anchor $e$ are
\[
\Pi(e)
=
\left\{
(e,t)
\;\middle|\;
\mathcal{N}_{1,t}(e)\neq\emptyset
\right\}.
\]
Each trajectory therefore represents a first-hop entity type that is
reachable from the candidate anchor.

Because entity types can be broad, their labels alone may provide
insufficient information about the content available along a trajectory.
We therefore introduce a \emph{preview} mechanism that retrieves the
$k_p$ first-hop relationships most semantically relevant to the query for
each trajectory. These previews provide concrete examples of relationships
matching the trajectory's first-hop structure and help the LLM determine
which trajectories to prioritize.
\paragraph{Trajectory Ranking.}
After trajectory discovery, the LLM globally ranks the candidate anchors
according to their relevance to the query. Given up to $k_a$ candidate
anchors
\[
\mathcal{A}=\{e_1,\ldots,e_m\},
\qquad m\leq k_a,
\]
it assigns each anchor a unique global rank
\[
\alpha:\mathcal{A}\rightarrow\{1,\ldots,m\},
\]
where $\alpha(e)=1$ denotes the highest-priority anchor and
$\alpha(e)=m$ denotes the lowest-priority anchor among the $m$ candidates.

For each anchor $e$, a separate LLM call then reads the query, the global
graph overview, the available entity types, and the relationship previews
associated with its trajectories
\[
\Pi(e)=\{\pi_1,\ldots,\pi_p\}.
\]
It assigns each trajectory a unique local rank
\[
\rho_e:\Pi(e)\rightarrow\{1,\ldots,p\},
\]
where $\rho_e(\pi)=1$ denotes the most promising trajectory and
$\rho_e(\pi)=p$ denotes the least promising trajectory for that anchor.

Each anchor--trajectory pair is therefore prioritized using
\[
\operatorname{Priority}(e,\pi)
=
\bigl(\alpha(e),\rho_e(\pi)\bigr),
\]
with global anchor rank considered first and local trajectory rank second.
\dotrag{} retains the top $k_\pi$ trajectories for each anchor, orders the
remaining pairs by this priority, and sends the top $k_\tau$ pairs to
relational path search. Their priorities are later inherited by the paths
recovered from them.

\paragraph{Relational Path Search.}
Relational path search begins from the top $k_\tau$ anchor--trajectory
pairs retained by the preceding stage. For each retained first-hop
trajectory $(e,t_1)$, \dotrag{} uses the reachability map to enumerate the
entity types that are structurally available as second-hop continuations.
These options are computed over the complete first-hop type frontier and
therefore represent possible continuations rather than guaranteeing that
each will remain reachable after first-hop semantic filtering.

The LLM generates a semantic query for $t_1$ and selects one or more
continuation types, generating a corresponding semantic query for each
selection. In Figure~\ref{fig:dotrag-retrieval}, it searches for
``Elon companies'' over the first-hop organizations and then
searches for ``Elon company products'' over the product entities
structurally reachable two hops from \textit{Elon Musk}.

At each hop, \dotrag{} uses the reachability map to progressively constrain
the search to relationships that extend the current frontier toward the
selected entity type. Let $\mathcal{F}_{h-1}$ contain the endpoints retained
from the previous hop, beginning with $\mathcal{F}_0=\{e\}$ for anchor $e$.
For each selected type and semantic-query pair $(t_h,q_h)$, the retained
relationships are
\[
\operatorname{Next}_h(\mathcal{F}_{h-1})
=
\underset{
\substack{
r=(u,v)\in\mathcal{R}\\
u\in\mathcal{F}_{h-1}\\
v\in\mathcal{N}_{h,t_h}(e)
}
}{
\operatorname{TopK}_{k_s}
}
\operatorname{sim}(q_h,d_r).
\]
The function retains the $k_s$ candidate relationships whose descriptions
$d_r$ are most similar to $q_h$. Each retained relationship extends the
path ending at $u$, and its endpoint becomes part of the frontier for the
next hop. Consequently, a continuation type that is structurally available
before first-hop retrieval contributes no evidence if it is no longer
reachable from the retained frontier. At subsequent hops, the available
continuation types are recomputed from the latest retained frontier.

In Figure~\ref{fig:dotrag-retrieval}, the first-hop search is restricted to
organizations connected to \textit{Elon Musk}. The retrieved organizations
then form the frontier for the product search, which is restricted to
products reachable through them. This produces connected paths such as
\textit{Elon Musk $\rightarrow$ Tesla $\rightarrow$ EVs} and
\textit{Elon Musk $\rightarrow$ SpaceX $\rightarrow$ Rockets}.

After each search round, the LLM reviews the evidence recovered at every
available depth and the entity types available at the next hop. Its
operational choices are
\[
\operatorname{Decision}_h
\in
\left\{
\begin{array}{l}
\textsc{Stop}(\mathcal{H}_{\mathrm{preferred}}),\\
\textsc{Continue}(t_{h+1},q_{h+1})
\end{array}
\right\}.
\]
When stopping, $\mathcal{H}_{\mathrm{preferred}}$ records the completed hop
depths that the LLM considers most relevant to the query. When continuing,
the LLM selects an available next-hop type $t_{h+1}$ and generates its
semantic query $q_{h+1}$. If the search reaches the maximum depth $H$, the
LLM performs a final evidence review and identifies the completed hop
depths that should receive priority before retrieval terminates.

In the figure, the two-hop results suggest that further evidence about
\textsc{Impact} may be useful. The LLM can therefore select
\textsc{Impact}, generate another query, and extend the search to produce
paths such as
\textit{Elon Musk $\rightarrow$ Tesla $\rightarrow$ EVs
$\rightarrow$ Emissions}.

\paragraph{Relational Path Ranking.}
Once relational path search concludes, \dotrag{} combines the paths
recovered from all surviving trajectories and removes exact duplicates.
Paths are ordered first by the priority inherited from their originating
trajectory and then by the evidence preferences produced during retrieval.
For a path $p=(r_1,\ldots,r_h)$, its cumulative semantic relevance is
\[
s(p)
=
\sum_{j=1}^{h}
\operatorname{sim}(q_j,d_{r_j}),
\]
where $q_j$ is the semantic query used at hop $j$ and $d_{r_j}$ is the
description of the relationship retrieved at that hop. This score orders
paths that share the same trajectory and evidence priorities.

The ranked paths are added to the final context until the token budget is
reached. Entities and relationships already introduced by a higher-ranked
path are not counted again, allowing overlapping paths to share evidence
without unnecessarily consuming the budget. Evidence from non-preferred
hop depths remains eligible when space is available.

%% file: sections/experiments.tex
\section{Experiments}
We evaluate DotRAG on relational retrieval benchmarks spanning varying graph scales and query complexities, addressing four research questions: (RQ1) How effectively does DotRAG retrieve relevant evidence across different reasoning depths and graph scales? (RQ2) How do retrieval improvements translate into downstream answer quality? (RQ3) How do DotRAG's components and backbone LLM affect retrieval performance? (RQ4) What are DotRAG's computational costs and scalability trade-offs?

\subsection{Datasets}
Our evaluation uses MetaQA, STaRK-MAG, and STaRK-Prime, which provide natural-language queries, densely connected knowledge graphs, and ground-truth answer entities. MetaQA is a movie knowledge graph connecting movies, actors, directors, years, ratings, and tags through relationships such as acting, directing, and release information. STaRK-MAG is an academic knowledge graph connecting research papers, authors, institutions, and fields of study through authorship, citation, and affiliation relationships. STaRK-Prime is a biomedical knowledge graph connecting diseases, drugs, genes, proteins, and biological pathways through biomedical relationships. We adapt MetaQA's complete knowledge graph to our GraphRAG setting, while evaluating STaRK-MAG and STaRK-Prime using subgraphs of their respective knowledge graphs. We provide dataset statistics, preprocessing details, and graph statistics in appendix~\ref{app:benchmarks}, appendix ~\ref{app:preprocess}, and Table ~\ref{tab:graph-statistics}, respectively.

\subsection{Evaluation Metrics}
We evaluate retrieval across four dimensions: retrieval accuracy, structural relevance, computational efficiency, and downstream answer quality. To measure retrieval accuracy, we use Recall, Precision, and F1 to assess how well retrieved entities match ground-truth answers. The MetaQA and STaRK benchmarks provide anchor entities, ground-truth answer entities, and annotated hop distances for each question, but do not annotate ground-truth relational paths. To avoid penalizing path-based retrieval algorithms for potentially relevant intermediate entities, we exclude non-answer entities lying on anchor-to-answer paths within each question’s specified hop distance from retrieval precision calculations. To assess structural relevance, we introduce Golden Proximity (GP), which measures the proportion of structurally relevant intermediate entities recovered during retrieval. This metric proxies the retrieval algorithm’s effectiveness at exploring structurally relevant graph regions by measuring its coverage of intermediate entities along anchor-to-answer paths. To assess computational efficiency, we also track the number of LLM calls made during retrieval for each query. Finally, we evaluate downstream answer quality using Answer Recall and F1, comparing the LLM's generated answers against the complete ground-truth answer set. We provide formal definitions in Appendix~\ref{app:evaluation-metrics}.

\subsection{Baselines}
Following \citet{guo-etal-2025-lightrag, chen2026pathrag}, we use GPT-4o-mini as the backbone LLM for all methods, with a uniform retrieval budget of 10,000 tokens. We select representative baselines from each retrieval paradigm: BM25 and vector similarity search (VSS) for Retrieve-Then-Generate, LightRAG and PathRAG for Reason-Retrieve-Generate, and HopRAG for Reason-In-Retrieval-Then-Generate. Baseline configurations are provided in Appendix~\ref{app:baseline-settings}.

\subsection{Retrieval Effectiveness (RQ1)}
As shown in Table~\ref{tab:entity-retrieval-results} and our expanded Table ~\ref{tab:metaqa-retrieval-results} in our appendix, DotRAG achieves the highest F1 across MetaQA hop settings and substantially improves Recall on multi-hop queries. On single-hop queries, most methods achieve high Recall, while reasoning-in-retrieval methods attain higher F1, suggesting greater selectivity in retrieved evidence. However, on two-hop queries, all baselines experience substantial declines in Recall, whereas DotRAG maintains its performance. Our Golden Proximity (GP) evaluation dimension reveals that although BM25, VSS, and LightRAG successfully retrieve entities within the intended anchor–answer boundary, they struggle to recover the actual answers. In contrast, DotRAG effectively translates focused graph exploration into higher Recall and F1. 

Additionally, on three-hop queries, identifying relevant graph regions becomes increasingly difficult, yet DotRAG achieves the highest golden proximity scores, Recall, and F1. Moreover, DotRAG consistently outperforms HopRAG across all hop settings while requiring substantially fewer LLM calls, demonstrating more effective and efficient retrieval-time reasoning.

Additionally, we evaluate DotRAG and the baselines on STaRK-MAG and STaRK-Prime, which provide more realistic and heterogeneous graph settings. Similar to MetaQA, reasoning-in-retrieval methods achieve the strongest F1 scores, with differences becoming more pronounced on two-hop queries. On STaRK-MAG, DotRAG achieves the highest two-hop Recall and F1. Although BM25 achieves a higher golden proximity score, DotRAG is more effective at translating relevant graph exploration into answer retrieval.

STaRK-Prime presents a greater challenge across all methods. Unlike STaRK-MAG, which focuses on academic publications, STaRK-Prime contains biomedical knowledge with specialized terminology, diverse entity types, and complex relational queries. Retrieval performance declines substantially, with VSS achieving the highest Recall and DotRAG achieving the highest two-hop F1. These results suggest that accurately interpreting domain-specific relational requirements may limit retrieval effectiveness, a dependency we investigate further in our backbone analysis.

Notably, PathRAG encounters prohibitive computational expense on both STaRK datasets, exceeding 20 minutes per query. This behavior stems from its enumeration of candidate relational paths, whose number can grow combinatorially in densely connected graphs. This limitation is one of the motivating examples for DotRAG's use of precomputed reachability information to constrain graph exploration without expensive online path enumeration.

\input{tables/main_result}

\subsection{Downstream Answer Quality (RQ2)}
Although a retrieval algorithm may retrieve more correct answer entities, we argue the LLM may not necessarily recognize or include them in its generated response. We therefore measure downstream performance to quantify how effectively the LLM utilizes retrieved information to produce correct answers. For this experiment, we select the best-performing retrieval algorithm from each paradigm: VSS, LightRAG, and DotRAG. We then provide our backbone LLM with the retrieved context and ask it to return a list of answer entities, which we use to compute answer-level recall, precision, and F1. We compare these scores with the retrieval algorithm’s raw recall, precision, and F1 to assess how much the LLM recognized the retrieved answers. 
\input{tables/answer_analysis}
Across all settings, we find that DotRAG consistently preserves more of its retrieval advantage in downstream answer generation compared to the baselines. On MetaQA 1-hop, STaRK-MAG 1-hop, and STaRK-Prime 1-hop, VSS achieves the highest retrieval recall, yet its answer-level recall drops to 67.08\%, 63.92\%, and 37.35\%, respectively. This suggests that retrieving more correct entities does not necessarily translate into better answer generation, as irrelevant or distracting surrounding context may hinder the LLM's ability to identify correct answers. Thus, high retrieval recall does not guarantee high answer recall. Another notable finding is that answer-level F1 substantially exceeds retrieval F1 across all settings, suggesting that the LLM effectively filters out irrelevant retrieved entities when generating its answers, albeit at the cost of reduced recall.

\subsection{Ablation study (RQ3)}
We conduct five ablation experiments to assess the contributions of DotRAG's key design components in Table ~\ref{tab:dotrag-ablation}. First, we replace all LLM-guided retrieval decisions with random decisions, eliminating LLM calls during retrieval and effectively transforming DotRAG into a retrieve-then-reason approach. Second, we remove the graph-wide description to evaluate its role in grounding the LLM's retrieval decisions in the graph's knowledge. Third, we remove the entity and relationship samples provided before traversal decisions to examine their influence on the LLM's ability to select relevant routes. Fourth, we remove the graph-reachability constraints from semantic search to investigate whether constrained search reduces noise introduced by globally prominent but irrelevant entities. Finally, we evaluate DotRAG using different backbone LLMs to examine whether its retrieval effectiveness generalizes across models.

\paragraph{Random Choices.}
We replace all LLM-guided retrieval decisions with random choices, reuse the initial query instead of generating new search queries, and randomly determine when to terminate retrieval. Despite removing LLM reasoning entirely, DotRAG achieves competitive recall on MetaQA 1-hop and outperforms all baselines on MetaQA 2-hop in both recall and F1. Notably, it achieves 47.81\% recall and 21.05\% F1 on MetaQA 2-hop without making any LLM calls. These results suggest that DotRAG's underlying retrieval design is effective even without LLM guidance, as its graph traversal mechanisms can narrow the search space and expose relevant answer entities despite random decision-making. However, this effectiveness diminishes as retrieval complexity increases. On MetaQA 3-hop, recall and F1 drop sharply to 13.90\% and 3.09\%, respectively. Similarly, on STaRK-Prime, performance deteriorates from 1-hop to 2-hop queries. These results suggest that while DotRAG's underlying retrieval design can identify relevant graph regions without LLM guidance, reasoning becomes increasingly important for navigating longer, more complex retrieval paths.

\paragraph{No Global Context.}
During initial graph indexing, a graph-wide description is generated to provide a general overview of the graph and a comprehensive summary of each entity type, grounding the LLM in the graph's knowledge to support informed retrieval decisions. We find that removing the graph-wide context has a small and inconsistent impact on retrieval performance. Notably, its removal results in a nearly 7-percentage-point drop in recall on STaRK-Prime 2-hop, while producing only modest changes across the remaining settings. Although the graph-wide description provides the LLM with a broader understanding of the graph's structure and contents, these results suggest that local graph information plays a more substantial role in guiding retrieval decisions than the global graph description.

\paragraph{No Preview.}
Before the LLM ranks anchor–trajectory pairs or generates follow-up search queries, each branch provides samples of relationship descriptions semantically similar to the current or previous query to support informed graph exploration. In this experiment, we replace all previews with the text "No Preview" to examine their influence on local retrieval decisions. Our results suggest that the effectiveness of previews is dataset-dependent. On STaRK-Prime, removing previews reduces recall by approximately 8–9 percentage points across both hop settings, with F1 also declining. This suggests that entity types and biomedical terminology alone may provide insufficient context for the LLM to make informed traversal decisions. On MetaQA and STaRK-MAG, removing previews reduces 1-hop recall by 3.39 and 8.57 percentage points, respectively. However, 2-hop recall improves substantially on STaRK-MAG, while MetaQA exhibits a slight improvement at 3 hops. These findings suggest that previews can help guide local retrieval decisions, but their effectiveness depends on the characteristics of the underlying knowledge graph and the retrieval depth.

\paragraph{No Constrained Search.}
In this experiment, we remove the constrained semantic search component from DotRAG. In our standard design, constrained search restricts retrieval candidates to entities of a specified type and hop distance, preventing semantically similar but structurally irrelevant entities from being retrieved. Without these constraints, semantic search operates over the global embedding space, where relevant candidates compete with entities across the entire graph. We also remove the requirement that newly retrieved entities must be reachable from the preceding retrieval trajectory. Our results show that removing constrained search substantially degrades retrieval performance across all datasets and hop settings. Notably, on MetaQA 2-hop, recall drops from 97.78\% to 36.18\%, while F1 decreases from 51.27\% to 3.45\%. These findings highlight constrained semantic search as a critical component of DotRAG, demonstrating the importance of incorporating graph structure into semantic retrieval to prevent irrelevant entities from dominating search results.

\paragraph{Model Sensitivity.}
We further evaluate DotRAG using GPT-4.1 and Mistral 3.2 24B-Instruct as alternative backbone LLMs to investigate how retrieval performance varies across models and whether DotRAG's effectiveness depends on the reasoning capabilities of the underlying LLM in Table ~\ref{tab:dotrag-model-scaling}. We find that GPT-4.1 generally achieves the highest recall and F1, suggesting that DotRAG benefits from stronger backbone models. Interestingly, Mistral also performs competitively in most evaluated settings, but its performance can deteriorate sharply as reasoning complexity increases, approaching that of random retrieval decisions. These findings suggest that while DotRAG generalizes across backbone models, its effectiveness on more complex retrieval tasks remains sensitive to the reasoning capabilities of the underlying LLM.

\subsection{Scalability and Cost Analysis  (RQ4)}
To evaluate the computational efficiency and scalability of DotRAG, we compare its retrieval cost against the baselines, including the number of LLM calls, latency, and associated token expenses in Table ~\ref{tab:retrieval-latency-results} and \ref{tab:retrieval-efficiency-results}. We also examine DotRAG's performance across knowledge graphs of varying sizes and structural complexity.

Our results show that DotRAG is substantially more computationally efficient than HopRAG, with both methods following the reason-in-retrieval-then-generate paradigm. For instance, on STaRK-MAG, DotRAG costs \$3.99 per 1,000 queries compared to \$11.21 for HopRAG, representing approximately a 64\% reduction in cost. Similar trends are observed across the remaining datasets, demonstrating that DotRAG achieves lower computational costs while incorporating LLM reasoning during retrieval.

Furthermore, we evaluate DotRAG across knowledge graphs containing approximately 40K--89K entities and 110K--487K relationships. Despite the size and structural complexity of these graphs, DotRAG maintains strong retrieval effectiveness while requiring relatively few LLM calls per query. These results demonstrate DotRAG's ability to operate efficiently on large, densely connected knowledge graphs without incurring excessive computational overhead.

We also observe that DotRAG achieves relatively low retrieval latency for a reason-in-retrieval-then-generate method. On STaRK-MAG, HopRAG requires an average of 156.87 seconds per query, compared to just 12.91 seconds for DotRAG. Despite incorporating LLM reasoning during retrieval, DotRAG maintains latency within the same order of magnitude as LightRAG and PathRAG on MetaQA. Furthermore, DotRAG remains practical on denser knowledge graphs, where PathRAG exceeds 20 minutes per query on both STaRK datasets. This suggests that DotRAG's retrieval design avoids the computational explosion associated with extensive online graph traversal, allowing it to incorporate retrieval-time reasoning while maintaining manageable computational costs.

\section{Conclusion}
In this work, we introduced DotRAG, a training-free GraphRAG retrieval algorithm that incorporates LLM reasoning directly into graph retrieval. DotRAG combines a global graph overview with precomputed reachability information to guide the exploration of task-relevant relational paths while constraining semantic search to structurally relevant candidates. Experiments on MetaQA and STaRK demonstrate substantial improvements in multi-hop retrieval effectiveness and downstream answer quality compared to existing retrieval paradigms. Our results further show that DotRAG achieves these gains with significantly lower computational overhead than HopRAG. Overall, these findings highlight the importance of combining LLM-guided retrieval decisions with graph-structural constraints for effective and efficient relational retrieval.

\label{sec:experiments}

%% file: tables/main_result.tex
\begin{table}[t]
\centering
\small
\renewcommand{\arraystretch}{1.08}

\resizebox{\linewidth}{!}{
\begin{tabular*}{\columnwidth}
{@{\extracolsep{\fill}} l cc ccc @{}}
\toprule
\textbf{Method}
& \multicolumn{2}{c}{\textbf{1-hop}}
& \multicolumn{3}{c}{\textbf{2-hop}} \\
\cmidrule(lr){2-3}
\cmidrule(lr){4-6}
& \textbf{Recall} & \textbf{F1}
& \textbf{Recall} & \textbf{F1} & \textbf{GP} \\
\midrule

\multicolumn{6}{l}{\cellcolor[HTML]{EFEFEF}\textit{MetaQA}} \\
\quad BM25
& 97.32 & 4.44
& 17.01 & 0.72 & \textbf{99.44} \\
\quad VSS
& \textbf{99.02} & 4.26
& \underline{34.74} & 2.08 & \underline{96.59} \\
\quad LightRAG
& 95.77 & 2.04
& 34.22 & 1.74 & 91.89 \\
\quad PathRAG
& 66.20 & 1.86
& 33.15 & 1.17 & 46.67 \\
\quad HopRAG (20\%)
& 83.25 & \underline{12.69}
& 31.82 & \underline{6.33} & 52.57 \\
\quad \textbf{\dotrag{}\textsubscript{ours}}
& \underline{97.56} & \textbf{14.39}
& \textbf{97.78} & \textbf{51.27} & 96.41 \\
\midrule

\multicolumn{6}{l}{\cellcolor[HTML]{EFEFEF}\textit{STaRK-MAG}}  \\
\quad BM25
& \textbf{95.77} & 10.74
& 63.84 & 3.96 & \textbf{92.44} \\
\quad VSS
& \underline{93.47} & 10.31
& \underline{77.75} & 4.98 & 63.17 \\
\quad LightRAG
& 75.64 & 8.20
& 67.45 & 4.32 & 49.54 \\
\quad PathRAG
& PCE & PCE
& PCE & PCE & PCE \\
\quad HopRAG (30\%)
& 88.78 & \textbf{19.93}
& 64.37 & \underline{11.85} & 14.68 \\
\quad \textbf{\dotrag{}\textsubscript{ours}}
& 89.33 & \underline{18.02}
& \textbf{85.07} & \textbf{15.75} & \underline{80.73} \\
\midrule


\multicolumn{6}{l}{\cellcolor[HTML]{EFEFEF}\textit{STaRK-Prime}} \\
\quad BM25
& 67.25 & 4.81
& 38.48 & 2.68 & 19.80 \\
\quad VSS
& \textbf{78.26} & 8.49
& \textbf{54.24} & 4.27 & \underline{22.52} \\
\quad LightRAG
& 69.76 & 5.57
& 44.55 & 3.05 & \textbf{25.12} \\
\quad PathRAG
& PCE & PCE
& PCE & PCE & PCE \\
\quad HopRAG (30\%)
& \underline{77.13} & \textbf{18.94}
& \underline{53.38} & \underline{10.50} & 20.75 \\
\quad \textbf{\dotrag{}\textsubscript{ours}}
& 56.71 & \underline{9.68}
& 48.52 & \textbf{12.23} & 21.96 \\
\bottomrule
\end{tabular*}}

\caption{Retrieval recall, F1, and GP on MetaQA, STaRK-MAG, and STaRK-Prime.
We evaluate each method with a retrieval budget of 10,000 tokens on gpt-4o-mini.
Best results per dataset are in \textbf{bold} and second-best are \underline{underlined}. 
HopRAG results are averaged over 100 questions per dataset due to its slow speed. PCE denotes Prohibitive Computational Expense.}
\label{tab:entity-retrieval-results}
\end{table}

%% file: tables/answer_analysis.tex
\begin{table*}[t]
\centering
\small
\renewcommand{\arraystretch}{1.08}
\setlength{\tabcolsep}{4pt}

\resizebox{\textwidth}{!}{%
\begin{tabular}{l cccc cccc cccc}
\toprule
\textbf{Dataset}
& \multicolumn{4}{c}{\textbf{\dotrag{}}}
& \multicolumn{4}{c}{\textbf{VSS}}
& \multicolumn{4}{c}{\textbf{LightRAG}} \\

\cmidrule(lr){2-5}
\cmidrule(lr){6-9}
\cmidrule(lr){10-13}

& \textbf{Ret. R} & \textbf{Ret. F1}
& \textbf{Ans. R} & \textbf{Ans. F1}
& \textbf{Ret. R} & \textbf{Ret. F1}
& \textbf{Ans. R} & \textbf{Ans. F1}
& \textbf{Ret. R} & \textbf{Ret. F1}
& \textbf{Ans. R} & \textbf{Ans. F1} \\
\midrule

\multicolumn{13}{l}{\textit{MetaQA}} \\

\quad 1-hop
& 97.56 & 14.39 & 86.11 & 82.52
& 99.02 & 4.26 & 67.08 & 63.25
& 95.77 & 2.04 & 74.87 & 69.14 \\

\quad 2-hop
& 97.78 & 51.27 & 85.94 & 83.94
& 34.74 & 2.08 & 19.78 & 19.96
& 34.22 & 1.74 & 35.24 & 30.48 \\

\quad 3-hop
& 63.48 & 21.19 & 44.70 & 45.46
& 38.15 & 5.15 & 14.16 & 13.40
& 37.31 & 4.06 & 19.21 & 19.19 \\

\hdashline

\multicolumn{13}{l}{\textit{STaRK-MAG}} \\

\quad 1-hop
& 89.33 & 18.02 & 71.69 & 65.86
& 93.47 & 10.31 & 63.92 & 58.74
& 75.64 & 8.20 & 34.51 & 35.73 \\

\quad 2-hop
& 85.07 & 15.75 & 68.91 & 61.07
& 77.75 & 4.98 & 36.36 & 29.16
& 67.45 & 4.32 & 25.23 & 23.41 \\

\hdashline

\multicolumn{13}{l}{\textit{STaRK-Prime}} \\

\quad 1-hop
& 56.71 & 9.68 & 39.80 & 35.59
& 78.26 & 8.49 & 37.35 & 32.23
& 69.76 & 5.57 & 24.25 & 22.30 \\

\quad 2-hop
& 48.52 & 12.23 & 37.55 & 33.50
& 54.24 & 4.27 & 26.98 & 23.25
& 44.55 & 3.05 & 18.68 & 17.27 \\

\bottomrule
\end{tabular}%
}

\caption{Retrieval and answer quality across MetaQA,
STaRK-MAG, and STaRK-Prime. Retrieval recall (Ret.\ R)
and F1 (Ret.\ F1) are the original retrieval results for
each method against the complete gold answer set. Answer
recall (Ans.\ R) and F1 (Ans.\ F1) evaluate the entities
subsequently identified by GPT-4o-mini against the same
complete gold answer set. Entities designated as intermediates,
but not answers, are excluded from scoring.}
\label{tab:answer-quality}
\end{table*}

%% file: sections/appendix.tex
\appendix

\section{Benchmark Selection}
GraphRAG evaluation is difficult to reproduce and interpret because several
system components often change simultaneously. Methods commonly construct
their own graphs from the same corpus, apply different retrieval algorithms,
serialize the retrieved evidence differently, and use method-specific
prompts for answer generation. LLM-based graph construction may also omit
or alter key entities and relationships across runs. Consequently, observed
answer-quality differences cannot be attributed cleanly to retrieval: they
may instead arise from graph extraction, evidence presentation, prompting,
or generation.

We therefore evaluate retrieval directly over pre-generated graphs. MetaQA and
STaRK allow every method to access identical entities, relationships, and
descriptions, while their relational structure supports deterministic
evaluation at increasing graph-hop distances. By holding graph construction,
evidence representation, retrieval budget, and generation model constant,
our evaluation isolates differences produced by the retrieval algorithms
themselves and yields conclusions that are more reproducible and
generalizable across future GraphRAG systems.
We select MetaQA and STaRK because their fixed relational structures allow
every method to retrieve from the same graph and enable deterministic
evaluation of answer and intermediate entities. This isolates retrieval
from graph construction and measures performance as the number of required
graph relationships increases. We preserve each benchmark's graph,
questions, and answers while adding textual descriptions required by
GraphRAG retrievers.

\section{Benchmark Overview}
\label{app:benchmarks}
\paragraph{STaRK \citep{wu2024stark}.}
STaRK-MAG contains 1,872,968 entities and 39,802,116 relationships, while
STaRK-Prime contains 129,375 entities and 8,100,498 relationships. Each
templated question identifies an anchor and its ground-truth answer
entities. We construct the evaluation graphs using 500 MAG questions and
1,000 Prime questions, including their anchors, answers, and connecting
entities and relationships. We further add randomly sampled graph elements
as controlled noise. Using more construction questions than evaluation
questions increases graph coverage and prevents retrieval from being
trivially restricted to the tested queries. The resulting MAG and Prime
graphs contain 88,815/486,616 and 54,148/399,657 entities/relationships,
respectively. We evaluate 345 questions from each graph.

\paragraph{MetaQA \citep{MetaQA}.}
MetaQA provides a pre-generated movie knowledge graph containing more than 40,000
entities and 100,000 relationships. Each question specifies a source entity
and ground-truth answer entities at a designated graph-hop distance. We
evaluate 500 questions from each of its one-, two-, and three-hop subsets,
totaling 1,500 questions.

\section{Preprocessing Details}
\label{app:preprocess}
STaRK provides entity names, descriptions, types, and relationship types.
We represent each entity using its provided name, description, and type,
and use the benchmark relation types as graph edges. For MetaQA, whose
entities originate from WikiMovies, we obtain entity descriptions from the
first passage of the corresponding Wikipedia article to introduce
realistic textual noise. Its graph edges are constructed directly from the
benchmark relationships.

For each question, we compute the intermediate entity set by finding all
anchor-to-answer paths matching its specified hop distance and excluding
the anchor and answer endpoints. All methods operate over the same
preprocessed entities, descriptions, and relationships.
\section{Baseline Configurations}
\label{app:baseline-settings}
All methods use GPT-4o-mini, a 10,000-token retrieval budget, and exclude
source chunks to evaluate entity-level retrieval only. LightRAG and
PathRAG use LightRAG's BYOG index; HopRAG represents entity descriptions
as nodes and relationship descriptions as edges. BM25 and VSS retrieve
the top 40 entities and relationships. LightRAG uses hybrid retrieval with
$k=40$ and allocates 2,000 and 6,000 tokens to entities and relationships,
respectively. PathRAG uses $k=60$ with 5,000 tokens each for its low- and
high-level contexts. HopRAG uses $k=20$ and a maximum depth of three hops.

For \dotrag{}, we set the seed-search cutoff to $k=40$, the maximum number
of candidate anchors to $k_a=5$, and the preview size to $k_p=3$.
Trajectory ranking retains $k_\pi=1$ trajectory per anchor and sends the
top $k_\tau=3$ anchor--trajectory pairs to relational path search. Each hop
retrieves up to $k_s=40$ relationships with a minimum similarity of $0.30$,
and the maximum search depth is $H=3$.
\subsection{Evaluation Metric Definitions}
\label{app:evaluation-metrics}
Let $\mathcal{R}$, $\mathcal{A}$, and $\mathcal{I}$ denote the retrieved,
answer, and intermediate entity sets, respectively, and define
$\mathcal{R}'=\mathcal{R}\setminus(\mathcal{I}\setminus\mathcal{A})$.
We report
\[
\begin{aligned}
R &=
\frac{|\mathcal{R}'\cap\mathcal{A}|}{|\mathcal{A}|},
&\qquad
P &=
\frac{|\mathcal{R}'\cap\mathcal{A}|}{|\mathcal{R}'|},\\
F1 &=
\frac{2PR}{P+R},
&
GP &=
\frac{|\mathcal{R}\cap\mathcal{I}|}{|\mathcal{I}|}.
\end{aligned}
\]
GP is N/A when $\mathcal{I}$ is empty. Answer-level metrics apply the same
$R$, $P$, and $F1$ definitions to entities selected by an LLM from the
retrieved context, comparing them against the complete answer set. Results
are macro-averaged. LLM calls count retrieval-time invocations separately,
including concurrent calls, and exclude answer generation. Latency excludes
initialization, warm-up, and answer generation.

\input{tables/ControlledHops}
\input{tables/RealWorldGraph}
\input{tables/latency}
\input{tables/ModelAnalysis}
\input{tables/Ablation}
\input{tables/CostAnalysis}
\input{tables/graph_statistics}
\input{tables/Bitmap}

%% file: tables/ControlledHops.tex
\begin{table*}[t]
\centering
\small
\renewcommand{\arraystretch}{1.08}

\begin{tabular*}{\textwidth}
{@{\extracolsep{\fill}} l cccc cccc cccc @{}}
\toprule
\textbf{Method}
& \multicolumn{12}{c}{\textbf{MetaQA}} \\

\cmidrule(lr){2-13}

& \multicolumn{4}{c}{\textbf{1-hop}}
& \multicolumn{4}{c}{\textbf{2-hop}}
& \multicolumn{4}{c}{\textbf{3-hop}} \\

\cmidrule(lr){2-5}
\cmidrule(lr){6-9}
\cmidrule(lr){10-13}

& \textbf{Recall} & \textbf{F1} & \textbf{Calls} & \textbf{GP}
& \textbf{Recall} & \textbf{F1} & \textbf{Calls} & \textbf{GP}
& \textbf{Recall} & \textbf{F1} & \textbf{Calls} & \textbf{GP} \\
\midrule

\multicolumn{13}{l}{\textit{Retrieve-Then-Generate}} \\
\quad BM25
& 97.32 & 4.44 & 0 & N/A
& 17.01 & 0.72 & 0 & 99.44
& 32.06 & 3.11 & 0 & 12.26 \\

\quad VSS
& 99.02 & 4.26 & 0 & N/A
& 34.74 & 2.08 & 0 & 96.59
& 38.15 & 5.15 & 0 & 13.03 \\

\hdashline

\multicolumn{13}{l}{\textit{Reason-Retrieve-Generate}} \\
\quad LightRAG
& 95.77 & 2.04 & 1 & N/A
& 34.22 & 1.74 & 1 & 91.89
& 37.31 & 4.06 & 1 & 16.35 \\

\quad PathRAG
& 66.20 & 1.86 & 1 & N/A
& 33.15 & 1.17 & 1 & 46.67
& 34.73 & 2.74 & 1 & 6.71 \\

\hdashline

\multicolumn{13}{l}{\textit{Reason-In-Retrieval-Then-Generate}} \\
\quad HopRAG (20\%)
& 83.25 & 12.69 & 61.53 & N/A
& 31.82 & 6.33 & 66.00 & 52.57
& 22.87 & 8.36 & 69.55 & 6.65 \\

\quad \textbf{\dotrag{}}
& 97.56 & 14.39 & 5.34 & N/A
& 97.78 & 51.27 & 4.62 & 96.41
& 63.48 & 21.19 & 5.65 & 24.84 \\

\bottomrule
\end{tabular*}

\caption{Retrieval recall (R), F1, and LLM calls on a random sample of 1,500 MetaQA questions, 500 per hop. Each method is evaluated with a retrieval budget of 10,000 tokens on gpt-4o-mini, filled exclusively with entities and relationships retrieved from the graph. The retrieved entities are compared against the gold entity set for each question. LLM calls refer to the average number of calls made during retrieval, excluding answer generation.}
\label{tab:metaqa-retrieval-results}
\end{table*}

%% file: tables/RealWorldGraph.tex
\begin{table*}[t]
\centering
\small
\renewcommand{\arraystretch}{1.08}
\setlength{\tabcolsep}{3.5pt}

\resizebox{\textwidth}{!}{%
\begin{tabular}{l cccc cccc cccc cccc}
\toprule
\textbf{Method}
& \multicolumn{16}{c}{\textbf{STaRK}} \\

\cmidrule(lr){2-17}

& \multicolumn{8}{c}{\textbf{STaRK-MAG}}
& \multicolumn{8}{c}{\textbf{STaRK-Prime}} \\

\cmidrule(lr){2-9}
\cmidrule(lr){10-17}

& \multicolumn{4}{c}{\textbf{1-hop}}
& \multicolumn{4}{c}{\textbf{2-hop}}
& \multicolumn{4}{c}{\textbf{1-hop}}
& \multicolumn{4}{c}{\textbf{2-hop}} \\

\cmidrule(lr){2-5}
\cmidrule(lr){6-9}
\cmidrule(lr){10-13}
\cmidrule(lr){14-17}

& \textbf{Recall} & \textbf{F1} & \textbf{Calls} & \textbf{GP}
& \textbf{Recall} & \textbf{F1} & \textbf{Calls} & \textbf{GP}
& \textbf{Recall} & \textbf{F1} & \textbf{Calls} & \textbf{GP}
& \textbf{Recall} & \textbf{F1} & \textbf{Calls} & \textbf{GP} \\
\midrule

\multicolumn{17}{l}{\textit{Retrieve-Then-Generate}} \\

\quad BM25
& 95.77 & 10.74 & 0 & N/A
& 63.84 & 3.96 & 0 & 92.44
& 67.25 & 4.81 & 0 & N/A
& 38.48 & 2.68 & 0 & 19.80 \\

\quad VSS
& 93.47 & 10.31 & 0 & N/A
& 77.75 & 4.98 & 0 & 63.17
& 78.26 & 8.49 & 0 & N/A
& 54.24 & 4.27 & 0 & 22.52 \\

\hdashline

\multicolumn{17}{l}{\textit{Reason-Retrieve-Generate}} \\

\quad LightRAG
& 75.64 & 8.20 & 1 & N/A
& 67.45 & 4.32 & 1 & 49.54
& 69.76 & 5.57 & 1 & N/A
& 44.55 & 3.05 & 1 & 25.12 \\

\quad PathRAG
& PCE & PCE & 1 & N/A
& PCE & PCE & 1 & PCE
& PCE & PCE & 1 & N/A
& PCE & PCE & 1 & PCE \\

\hdashline

\multicolumn{17}{l}{\textit{Reason-In-Retrieval-Then-Generate}} \\

\quad HopRAG (30\%)
& 88.78 & 19.93 & 62.85 & N/A
& 64.37 & 11.85 & 65.96 & 14.68
& 77.13 & 18.94 & 73.42 & N/A
& 53.38 & 10.50 & 75.28 & 20.75 \\

\quad \textbf{\dotrag{}}
& 89.33 & 18.02 & 7.84 & N/A
& 85.07 & 15.75 & 8.66 & 80.73
& 56.71 & 9.68 & 9.17 & N/A
& 48.52 & 12.23 & 8.94 & 21.96 \\

\bottomrule
\end{tabular}%
}

\caption{Retrieval recall (R), F1, and LLM calls on random
samples of STaRK-MAG and STaRK-Prime questions. Each method is
evaluated with a retrieval budget of 10,000 tokens on
gpt-4o-mini, filled exclusively with entities and relationships
retrieved from the graph. The retrieved entities are compared
Prohibitive Computational Expense.}
\label{tab:stark-entity-retrieval-results}
\end{table*}

%% file: tables/latency.tex
\begin{table*}[t]
\centering
\small
\renewcommand{\arraystretch}{1.08}

\begin{tabular*}{\textwidth}
{@{\extracolsep{\fill}} l c c c c c @{}}
\toprule
\textbf{Method}
& \multicolumn{3}{c}{\textbf{MetaQA}}
& \multicolumn{2}{c}{\textbf{STaRK}} \\

\cmidrule(lr){2-4}
\cmidrule(lr){5-6}

& \textbf{1-hop}
& \textbf{2-hop}
& \textbf{3-hop}
& \textbf{MAG}
& \textbf{Prime} \\

& \textbf{Time}
& \textbf{Time}
& \textbf{Time}
& \textbf{Time}
& \textbf{Time} \\
\midrule

\quad BM25
& 0.16 s & 0.17 s & 0.21 s & 0.96 s & 0.57 s \\

\quad VSS
& 0.28 s & 0.28 s & 0.27 s & 2.35 s & 1.78 s \\

\hdashline

\quad LightRAG
& 2.53 s & 2.71 s & 2.59 s & 4.38 s & 4.38 s \\

\quad PathRAG
& 4.65 s & 4.77 s & 4.88 s & $\geq$20 min & $\geq$20 min \\

\hdashline

\quad HopRAG
& 94.15 s & 82.16 s & 87.93 s & 156.87 s & 81.95 s \\

\quad \textbf{\dotrag{}}
& 7.44 s & 6.72 s & 8.43 s & 12.91 s & 11.40 s \\

\bottomrule
\end{tabular*}

\caption{Average retrieval latency per question across MetaQA and
STaRK. Latency measures retrieval only and excludes answer
generation. PathRAG queries on STaRK-MAG and STaRK-Prime exceeded
20 minutes.}
\label{tab:retrieval-latency-results}
\end{table*}

%% file: tables/ModelAnalysis.tex
\begin{table*}[tp]
\centering
\small
\renewcommand{\arraystretch}{1.08}
\setlength{\tabcolsep}{5pt}

\begin{tabular*}{\textwidth}
{@{\extracolsep{\fill}} l cccc cccc cccc @{}}
\toprule
\textbf{Dataset}
& \multicolumn{4}{c}{\textbf{GPT-4o-mini}}
& \multicolumn{4}{c}{\textbf{GPT-4.1}}
& \multicolumn{4}{c}{\textbf{Mistral 3.2 24B-Instruct}} \\

\cmidrule(lr){2-5}
\cmidrule(lr){6-9}
\cmidrule(lr){10-13}

& \textbf{R} & \textbf{F1} & \textbf{Calls} & \textbf{GP}
& \textbf{R} & \textbf{F1} & \textbf{Calls} & \textbf{GP}
& \textbf{R} & \textbf{F1} & \textbf{Calls} & \textbf{GP} \\
\midrule

\multicolumn{13}{l}{\textit{MetaQA}} \\

\quad 1-hop
& 97.56 & 14.39 & 5.34 & N/A
& \textbf{98.00} & \textbf{63.03} & 4.65 & N/A
& 97.50 & 14.15 & 6.23 & N/A \\

\quad 2-hop
& 97.78 & 51.27 & 4.62 & 96.41
& 98.12 & 52.59 & 4.54 & \textbf{97.09}
& \textbf{98.13} & \textbf{54.78} & 4.47 & 96.71 \\

\quad 3-hop
& 63.48 & 21.19 & 5.65 & 24.84
& \textbf{73.89} & \textbf{25.22} & 5.71 & 25.76
& 15.56 & 4.54 & 5.01 & \textbf{27.69} \\

\hdashline

\multicolumn{13}{l}{\textit{STaRK-MAG}} \\

\quad 1-hop
& 89.33 & 18.02 & 7.84 & N/A
& \textbf{95.85} & \textbf{22.19} & 6.90 & N/A
& -- & -- & -- & N/A \\

\quad 2-hop
& 85.07 & 15.75 & 8.66 & 80.73
& \textbf{96.77} & \textbf{16.19} & 8.31 & \textbf{85.46}
& -- & -- & -- & -- \\

\hdashline

\multicolumn{13}{l}{\textit{STaRK-Prime}} \\

\quad 1-hop
& 56.71 & 9.68 & 9.17 & N/A
& \textbf{65.68} & \textbf{13.71} & 9.32 & N/A
& -- & -- & -- & N/A \\

\quad 2-hop
& 48.52 & 12.23 & 8.94 & \textbf{21.96}
& \textbf{65.13} & \textbf{17.78} & 8.69 & 21.09
& -- & -- & -- & -- \\

\bottomrule
\end{tabular*}

\caption{Retrieval recall (R), F1, and average LLM calls
for \dotrag{} using different language models on MetaQA,
STaRK-MAG, and STaRK-Prime. All configurations use the
same graph index, retrieval algorithm, and retrieval budget
of 10,000 tokens. Retrieved entities are evaluated against
the gold entity set. LLM calls exclude answer generation.}
\label{tab:dotrag-model-scaling}
\end{table*}

%% file: tables/Ablation.tex
\begin{table*}[tp]
\centering
\small
\renewcommand{\arraystretch}{1.08}
\setlength{\tabcolsep}{4pt}

\resizebox{\textwidth}{!}{%
\begin{tabular}{l cccc cccc cccc cccc cccc}
\toprule
\textbf{Dataset}
& \multicolumn{4}{c}{\textbf{Original}}
& \multicolumn{4}{c}{\textbf{Random Choices}}
& \multicolumn{4}{c}{\textbf{No Global Context}}
& \multicolumn{4}{c}{\textbf{No Preview}}
& \multicolumn{4}{c}{\textbf{No Constrained Search}} \\

\cmidrule(lr){2-5}
\cmidrule(lr){6-9}
\cmidrule(lr){10-13}
\cmidrule(lr){14-17}
\cmidrule(lr){18-21}

& \textbf{Recall} & \textbf{F1} & \textbf{Calls} & \textbf{GP}
& \textbf{Recall} & \textbf{F1} & \textbf{Calls} & \textbf{GP}
& \textbf{Recall} & \textbf{F1} & \textbf{Calls} & \textbf{GP}
& \textbf{Recall} & \textbf{F1} & \textbf{Calls} & \textbf{GP}
& \textbf{Recall} & \textbf{F1} & \textbf{Calls} & \textbf{GP} \\
\midrule

\multicolumn{21}{l}{\textit{MetaQA}} \\

\quad 1-hop
& 97.56 & 14.39 & 5.34 & N/A
& 64.24 & 14.14 & 0 & N/A
& 96.96 & 14.56 & 5.33 & N/A
& 94.17 & 11.21 & 5.88 & N/A
& 69.26 & 5.89 & 5.32 & N/A \\

\quad 2-hop
& 97.78 & 51.27 & 4.62 & 96.41
& 47.81 & 21.05 & 0 & 90.35
& 97.82 & 50.35 & 4.68 & 96.28
& 96.44 & 36.68 & 5.05 & 94.99
& 36.18 & 3.45 & 5.00 & 18.01 \\

\quad 3-hop
& 63.48 & 21.19 & 5.65 & 24.84
& 13.90 & 3.09 & 0 & 19.81
& 62.80 & 20.84 & 5.59 & 24.90
& 64.46 & 22.29 & 5.72 & 23.60
& 37.45 & 5.97 & 5.44 & 10.39 \\

\hdashline

\multicolumn{21}{l}{\textit{STaRK-MAG}} \\

\quad 1-hop
& 89.33 & 18.02 & 7.84 & N/A
& 48.50 & 12.33 & 0 & N/A
& 88.72 & 19.82 & 7.47 & N/A
& 80.76 & 15.56 & 8.35 & N/A
& 80.69 & 15.49 & 7.73 & N/A \\

\quad 2-hop
& 85.07 & 15.75 & 8.66 & 80.73
& 56.31 & 10.74 & 0 & 54.86
& 87.25 & 14.82 & 8.43 & 79.64
& 94.68 & 16.33 & 8.87 & 83.64
& 62.35 & 7.15 & 8.59 & 1.09 \\

\hdashline

\multicolumn{21}{l}{\textit{STaRK-Prime}} \\

\quad 1-hop
& 56.71 & 9.68 & 9.17 & N/A
& 35.90 & 5.74 & 0 & N/A
& 52.96 & 10.41 & 9.10 & N/A
& 48.73 & 7.66 & 9.71 & N/A
& 49.54 & 6.59 & 9.06 & N/A \\

\quad 2-hop
& 48.52 & 12.23 & 8.94 & 21.96
& 14.44 & 3.15 & 0 & 14.91
& 41.64 & 11.61 & 8.71 & 18.74
& 39.62 & 8.31 & 8.98 & 17.91
& 36.40 & 4.73 & 8.73 & 13.72 \\

\bottomrule
\end{tabular}%
}

\caption{Ablation study of \dotrag{} across MetaQA,
STaRK-MAG, and STaRK-Prime. We compare the original
retrieval algorithm against four variants: random retrieval
choices, removal of the global graph overview, removal
of the preview mechanism, and removal of constrained search.
Performance is measured using retrieval recall, F1,
average LLM calls, and GP.}
\label{tab:dotrag-ablation}
\end{table*}

%% file: tables/CostAnalysis.tex
\begin{table*}[tp]
\centering
\scriptsize
\renewcommand{\arraystretch}{1.12}
\setlength{\tabcolsep}{2.5pt}

\resizebox{\textwidth}{!}{%
\begin{tabular}{l *{5}{rrr}}
\toprule
\textbf{Method}
& \multicolumn{9}{c}{\textbf{MetaQA}}
& \multicolumn{6}{c}{\textbf{STaRK}} \\

\cmidrule(lr){2-10}
\cmidrule(lr){11-16}

& \multicolumn{3}{c}{\textbf{1-hop}}
& \multicolumn{3}{c}{\textbf{2-hop}}
& \multicolumn{3}{c}{\textbf{3-hop}}
& \multicolumn{3}{c}{\textbf{MAG}}
& \multicolumn{3}{c}{\textbf{Prime}} \\

\cmidrule(lr){2-4}
\cmidrule(lr){5-7}
\cmidrule(lr){8-10}
\cmidrule(lr){11-13}
\cmidrule(lr){14-16}

& \textbf{Input} & \textbf{Output} & \textbf{USD/1K}
& \textbf{Input} & \textbf{Output} & \textbf{USD/1K}
& \textbf{Input} & \textbf{Output} & \textbf{USD/1K}
& \textbf{Input} & \textbf{Output} & \textbf{USD/1K}
& \textbf{Input} & \textbf{Output} & \textbf{USD/1K} \\
\midrule

\quad LightRAG
& 419.51 & 41.93 & \$0.09
& 422.75 & 43.46 & \$0.09
& 425.34 & 44.09 & \$0.09
& 440.57 & 57.27 & \$0.10
& 439.12 & 60.55 & \$0.10 \\

\quad PathRAG
& 387.51 & 41.21 & \$0.08
& 390.75 & 43.45 & \$0.08
& 393.34 & 44.46 & \$0.09
& PCE & PCE & PCE
& PCE & PCE & PCE \\

\hdashline

\quad HopRAG
& 61,326.25 & 543.51 & \$9.53
& 66,016.99 & 582.08 & \$10.25
& 69,851.60 & 608.65 & \$10.84
& 72,617.54 & 534.09 & \$11.21
& 87,014.53 & 621.31 & \$13.42 \\

\quad \textbf{\dotrag{}}
& 7,334.85 & 472.19 & \$1.38
& 5,437.07 & 375.42 & \$1.04
& 9,821.08 & 555.14 & \$1.81
& 23,190.08 & 858.17 & \$3.99
& 18,001.65 & 952.13 & \$3.27 \\

\bottomrule
\end{tabular}%
}

\caption{Average input and output tokens per question and
estimated cost (USD per 1,000 queries) across MetaQA and
STaRK. Costs are calculated using GPT-4o-mini pricing
(\$0.15/M input tokens; \$0.60/M output tokens).
PCE denotes Prohibitive Computational Expense. HopRAG results on
STaRK-MAG and STaRK-Prime are each averaged over 100 questions.}
\label{tab:retrieval-efficiency-results}
\end{table*}

%% file: tables/graph_statistics.tex
\begin{center}
\scriptsize
\renewcommand{\arraystretch}{1.05}
\setlength{\tabcolsep}{3pt}

\resizebox{\columnwidth}{!}{%
\begin{tabular}{lrrrrr}
\toprule
\textbf{Graph}
& \textbf{\# Entities}
& \textbf{\# Relationships}
& \textbf{Entity Tokens}
& \textbf{Relationship Tokens}
& \textbf{Database Size} \\
\midrule

MetaQA
& 40,151
& 109,605
& 5,182,672
& 10,465,503
& 5.63 GiB \\

STaRK-MAG
& 88,815
& 486,616
& 22,922,196
& 56,873,555
& 9.74 GiB \\

STaRK-Prime
& 54,148
& 399,657
& 9,594,981
& 40,662,194
& 6.76 GiB \\

\bottomrule
\end{tabular}%
}

\captionof{table}{Statistics of the graphs used in our evaluation.
Token totals are computed from the stored entity and
relationship descriptions. Database size includes graph
records, indexes, vector embeddings, and supporting metadata.}
\label{tab:graph-statistics}
\end{center}

\begin{center}
\small
\renewcommand{\arraystretch}{1.05}
\setlength{\tabcolsep}{5pt}

\resizebox{\columnwidth}{!}{%
\begin{tabular}{lrrrr}
\toprule
\textbf{Graph}
& \textbf{\# 1-hop}
& \textbf{\# 2-hop}
& \textbf{\# 3-hop}
& \textbf{\# Total} \\
\midrule

MetaQA
& 500
& 500
& 500
& 1,500 \\

STaRK-MAG
& 161
& 184
& N/A
& 345 \\

STaRK-Prime
& 145
& 200
& N/A
& 345 \\

\bottomrule
\end{tabular}%
}

\captionof{table}{Distribution of evaluation questions by relational hop count.}
\label{tab:hop-distribution}
\end{center}

%% file: tables/Bitmap.tex
\begin{center}
\scriptsize
\renewcommand{\arraystretch}{0.95}
\setlength{\tabcolsep}{2pt}

\resizebox{0.95\columnwidth}{!}{%
\begin{tabular}{llrrrrr}
\toprule
\textbf{Graph}
& \textbf{Variant}
& \textbf{Entities}
& \textbf{Relations}
& \textbf{Graph Size}
& \textbf{Build Time}
& \textbf{Index Size} \\
\midrule

MetaQA
& Full
& 40K
& 110K
& 23.56 MiB
& 29.91 s
& 388.67 MiB \\

\midrule

\multirow{2}{*}{STaRK-MAG}
& Sample
& 89K
& 487K
& 164.49 MiB
& 267.20 s
& 1.85 GiB \\

& Full
& 1.87M
& 39.8M
& 1,449.36 MiB
& --
& -- \\

\midrule

\multirow{2}{*}{STaRK-Prime}
& Sample
& 54K
& 400K
& 67.49 MiB
& 33.94 s
& 712.60 MiB \\

& Full
& 129K
& 8.1M
& 163.03 MiB
& --
& $\sim$4.1 GiB \\

\bottomrule
\end{tabular}%
}

\captionof{table}{Construction time and storage
requirements of exact three-hop reachability indexes
on sampled and full knowledge graphs. Graph size
refers to the serialized graph representation.
Entity and relationship counts are rounded.}
\label{tab:reachability-index-statistics}
\end{center}